# HD snapshot diffractive spectral imaging and inferencing

Apratim Majumder,[1]* Monjurul Meem, [1]† Fernando Gonzalez del Cueto,[2] Fernando Guevara-Vasquez,[2,3] Syed N. Qadri,[4] Freddie Santiago,[4] and Rajesh Menon[1,2]

[1]Dept. of Electrical & Computer Engineering, University of Utah, Salt Lake City, UT 84112, USA.
[2]Lumos Imaging Inc., Salt Lake City UT, 84103, USA.
[3]Dept. of Mathematics, University of Utah, Salt Lake City, UT 84112, USA.
[4]U.S. Naval Research Laboratory, Washington, DC 20036, USA.
[*] corresponding author, email: apratim.majumder@utah.edu
[†]Currently with Intel Corp., Hillsboro, OR 97124, USA.

**Abstract**
We present a novel high-definition (HD) snapshot diffractive spectral imaging system utilizing a diffractive filter array (DFA) to capture a single image that encodes both spatial and spectral information. This single diffractogram can be computationally reconstructed into a spectral image cube, providing a high-resolution representation of the scene across 25 spectral channels in the 440-800 nm range at 1304×744 spatial pixels (~1 MP). This unique approach offers numerous advantages including snapshot capture, a form of optical compression, flexible offline reconstruction, the ability to select the spectral basis after capture, and high light throughput due to the absence of lossy filters. We demonstrate a 30–50 nm spectral resolution and compared our reconstructed spectra against ground truth obtained by conventional spectrometers. Proof-of-concept experiments in diverse applications including biological tissue classification, food quality assessment, and simulated stellar photometry validate our system's capability to perform robust and accurate inference. These results establish the DFA-based imaging system as a versatile and powerful tool for advancing scientific and industrial imaging applications.

**Teaser**
Novel imaging system enables HD and 25-band spectral imaging via diffractive compressive filters and computational reconstruction.

## Introduction

Spectral information from a scene can significantly enhance performance in tasks such as target localization, classification, segmentation, and anomaly detection. This improvement arises because the interaction of light with matter produces wavelength-dependent signatures through reflection and absorption, offering valuable insights into material composition (*1–4*). However, conventional imagers typically produce either panchromatic images or aggregate data into few bands using filters, such as Bayer RGB cameras, which are designed to mimic human vision. Imaging spectroscopy addresses this gap by capturing the spatio-spectral information from a scene as 3D data cubes, where intensity is recorded as a function of space and spectrum (*5*, *6*). The spectral channels, ranging from tens of bands in multispectral to hundreds in hyperspectral, allow better material differentiation, benefiting fields such as medical imaging, (*7–9*), environmental monitoring (*10*), agriculture (*11*, *12*), geology (*13*), quality control (*14, 15*), astronomy (*16*), machine-vision and various other applications (*17–18*).

Despite these advantages, traditional hyperspectral imaging systems rely on scanning methods, compiling 2D images to construct the 3D cube, $I(x, y, \lambda)$, which can be expensive, bulky, and difficult to scale. 2D scans necessitate stationary scenes to avoid data smearing. The push-broom method disperses light via a grating or prism and scans relative to the subject, suitable for specific scenarios like airborne imaging *(19–21)*. Other techniques use filters to capture multiple shots over time *(22)*, with technologies like liquid-crystal and acousto-optic tunable filters adjusting the spectrum *(23, 24)*. Such 2D methods are limited for applications needing the entire image $I(x, y, \lambda)$ in one shot, risking motion blur and aberrations due to insufficient speed.

Snapshot methods to overcome these constraints have also been proposed. Coded apertures may offer high resolution, but suffer from absorption losses, sometimes bulky hardware, and limited spectral resolution *(25-27)*. Multi-aperture filters capture multiple spectral bands quickly, but also suffer from absorption losses, low spatial resolution, expensive fabrication, and often face alignment and calibration challenges *(28, 29)*. Fabry–Pérot resonators enhance portability, but suffer from low transmission, reduced spectral coverage, small field-of-view and high costs *(30, 31)*. Plasmonic approaches suffer from fabrication complexity and tend to be lossy *(32, 33)*. Speckle correlations and compressive sensing as well as other lensless-imaging techniques reduce measurement requirements, but require extensive calibration *(34)*. Metasurfaces enable precision and compact imaging, but struggle with high costs and manufacturing challenges, and tend to be very lossy *(35-39)*.

Previously, we explored an alternative paradigm to address these challenges by replacing a conventional Bayer filter with a transparent diffractive filter and then reconstructing the spectral information using computational methods *(40-43)*. In ref. *(42)*, we enhanced light sensitivity up to 3.12 times, while increasing spectral channels to 5 (from the conventional 3 in RGB). Subsequently, we demonstrated *diffractive-spectral imaging* of 25 bands at video-rates by placing a diffractive-filter array (DFA) in close proximity to an image sensor in a conventional camera *(43)*. The DFA microstructure causes the incoming image to be slightly diffracted when it reaches the sensor, producing wavelength-dependent spatial patterns in the recorded image. We call this monochrome image the *diffractogram*. By accurately modeling and measuring the response of a DFA-equipped optical system, we can computationally invert the process to reconstruct a substantive portion of the spatial and spectral content of the originating scene (Fig. 1A). In this work, we describe this methodology and demonstrate that spectral images reconstructed from diffractograms can be effectively used for some inference tasks across different fields, showcasing the versatility and potential of this approach.

Using this approach, we previously achieved 9.6 nm spectral resolution over a wavelength range of 430–718 nm, and even demonstrated a 30% spatial resolution enhancement by localization of the structured point-spread function (PSF) *(43)*. Furthermore, this approach allows computational trade-offs between spectral resolution and field of view without hardware changes, while maintaining high sensitivity by avoiding absorptive filters. However, all prior works were limited to a very small image of less than 50×50 pixels. In fact, most computational spectral imaging suffers from the problem of very small number of spatial pixels.

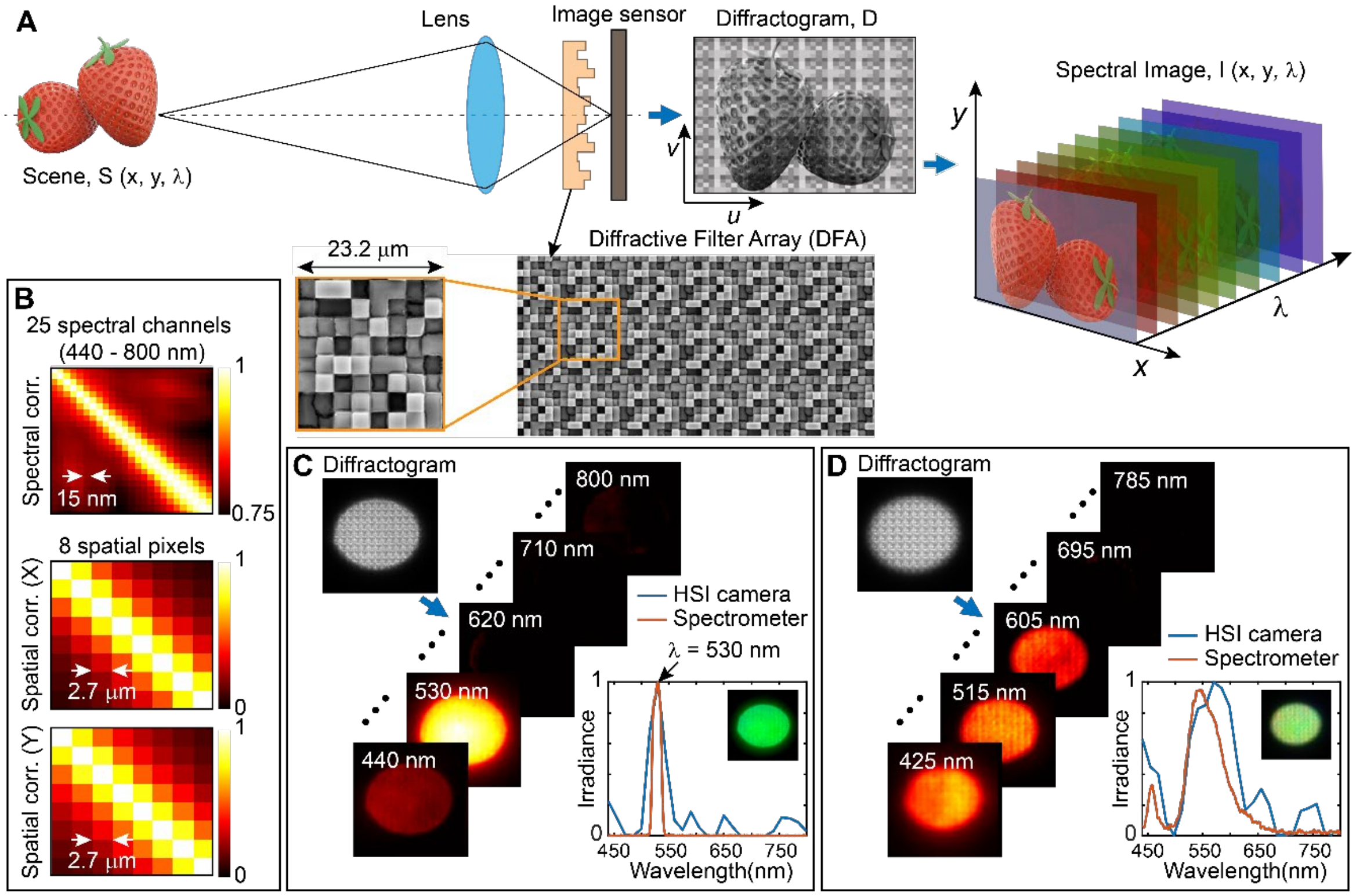

**Fig. 1: Placing a diffractive filter array (DFA) in close proximity to a CMOS image sensor enables diffractive-spectral imaging. (A)** The monochrome image sensor records a 2D diffractogram, which gives the 3D spatio-spectral image after computational reconstruction. Inset: scanning electron micrographs of the DFA master fabricated in Silicon. Additional photographs of the camera after assembly as well as the DFA are provided in Fig. S2. **(B)** Spectral and spatial (X and Y) responses of our computational camera. The responses are averaged over the orthogonal dimensions. **(C)** Example of a narrowband image (133 pixels sq.) of a white paper illuminated by narrowband light centered at 530nm (bandwidth =15nm). Only 5 of the 25 spectral channels are shown for clarity. Bottom-right inset shows average spectrum (orange curve is ground truth measured using a conventional spectrometer). **(D)** Example of a broadband image (122 pixels sq.) of a green Macbeth color chart illuminated by white light. All spectral channels and other examples are included in Fig. S6 – S8 of the Supplementary Information.

In this work, we solve this fundamental problem by extending diffractive-spectral imaging to HD spectral images (1304×744 spatial pixels, ~1MP) with 25 spectral channels in the 440–800 nm range, an increase of more than 2 orders of magnitude in spatio-spectral information. This is achieved without compromising imaging speed, which is only limited by the native frame rate of the (conventional) camera. Since spectral information is encoded within the grayscale image, we achieve a corresponding reduction in storage needs and data-transmission rates, which is crucial for remote sensing applications. Finally, we also apply standard inferencing to the reconstructed spectral images in order to: (1) classify lungs and trachea tissues in *ex vivo* chicken-lung images; (2) predict the freshness (age) of strawberries; and (3) generate Johnson/Cousins-filter responses for synthetic stellar objects.

## Working Principle

Our goal is to reconstruct a 3D spectral scene, $S(x, y, \lambda)$ from a 2D diffractogram, $D(u, v)$, as illustrated in Fig. 1(A). The forward problem can be modeled as:

$$D(u, v) = \int S(x, y, \lambda) A(u, v; x, y, \lambda)\, dxdyd\lambda. \tag{1}$$

Here, $A(u, v; x, y, \lambda)$ is the spatio-spectral point spread function (PSF), representing the response of the system to a point source located at $(x, y)$ and wavelength $\lambda$. This integral can be interpreted as a forward linear operator $F[A]$, mapping the spectral scene $S(x, y, \lambda)$ to the diffractogram $D(u, v)$. In general, the PSF is space variant so PSF acquisition for every value of $(x, y, \lambda)$ is impractical. Here we use a DFA that is $(p_1, p_2)$-periodic, which implies that the PSF of the system satisfies $A(u + n_1 p_1, v + n_2 p_2; x + n_1 p_1, y + n_2 p_2, \lambda) = A(u, v; x, y, \lambda)$ for any integers $n_1, n_2$. The PSF periodicity simplifies the PSF acquisition since we only need to take images for $x, y$ in a unit cell and $\lambda$ in the desired spectral discretization. We call this process *PSF calibration* and it is further detailed in the Materials and Methods section. Moreover, the periodicity of the PSF can be exploited to evaluate a discretization of $F[A]$ efficiently using the Discrete Fourier Transform. Since the goal is to invert the forward linear mapping $F[A]$, it is crucial to design $A$ to preserve as much information from the object $V$ as possible. The two critical design parameters are the DFA geometry and the gap between the DFA and the sensor. The choice of the gap is critical: a small gap results in a diffraction pattern that changes little with wavelength and a large gap introduces too much spatial mixing. Both extremes make the spectral reconstructions much harder. We optimize the gap and DFA geometry via a large stochastic search, aiming at making the simulated diffraction patterns corresponding to constant objects as different as possible across wavelengths, and respecting fabrication constraints resulting in an optimal design shown by the fabricated device in inset of Fig. 1(A), as well as in high-resolution optical and scanning electron micrographs in Supplementary Information Fig. S2.

The inverse problem can be formulated as (*44*):

$$\min_{V} \|F[A]V - D\|_2^2 + R(V; \theta). \tag{2}$$

Here, the first term represents the least-squares data-fitting functional, where $F[A]$ is the forward map in Equation (1). The second term is a regularization functional $R$, parametrized by $\theta$, which incorporates prior knowledge and constraints on $V$. Regularization is necessary to solve this optimization problem because the dimensionality of the unknown $V$ is, to fix ideas, $N$ times larger than that of the data $I$, where $N$ is the number of spectral bands. Without regularization, numerical solutions to the optimization problem tend to be highly oscillatory and can diverge severely from the ground truth (a manifestation of overfitting an erroneous model to noisy data). Regularization mitigates this by penalizing solutions deviating from prior knowledge. Here we used Total Variation (TV) regularization which penalizes noise and unnatural artifacts without smoothing sharp edges. TV regularization can be incorporated into the optimization problem through alternating methods (*45–48*).

While recovering a three-dimensional dataset from a two-dimensional projection may appear at odds with fundamental results of linear algebra, it is important to recognize that spectral images typically lie on a lower-dimensional manifold embedded in high-dimensional space. This can be understood from the inherent structure and compressibility of natural images and the very high correlation between spectral bands, as hyperspectral images are known to be well approximated locally by low-rank tensors *(49-52)*. Our reconstruction algorithm, by incorporating prior knowledge about the known structure of spectral images via regularization, attempts to recover the low-dimensional structures embedded within its 3D representation that can best explain the given data. In many cases, these recovered structures are sufficient to perform inference tasks accurately and robustly, as is supported by experimental results included here.

The periodic structure of the DFA enables the use of the measured PSF over a wide neighborhood around its location. However, since most camera lenses are not perfect, the intrinsic space variance of the camera lens introduces a long-range (but slowly varying) error to the PSF across the full frame of the sensor (field-of-view of the scene). We empirically determined that the PSF remains consistent for a distance up to 200 pixels from where it is measured. Therefore, in order to attain HD or full-frame imaging (1304×744 spatial pixels, × 25 spectral bands), we measured the PSF at ten different locations across the sensor frame (see Fig. S4). At each corresponding location in the scene, we placed a point source and recorded its diffractogram. Further details are provided in the Materials and Methods section. The spectral band (25 bands from 440 to 800 nm) of the point source as well as its location within one unit cell of the DFA (8×8 positions) was selectable, resulting in a measurement of the PSF, $A(u, v; x, y, \lambda)$. We can estimate the expected spatio-spectral performance of the reconstructions by cross-correlating the measured PSF in each spatio-spectral dimension. These cross-correlation maps averaged over the ten PSF measurements spanning the full-frame of the sensor are shown in Fig. 1(B). The spread of the diagonals of these maps offer an indication of the achievable spatio-spectral resolution (~one sensor pixel in space and 30-50 nm in spectrum), which was experimentally confirmed via spectral response functions. Furthermore, we performed extensive characterization of the spectral-reconstruction errors via standard experiments, two of which are summarized in Figs. 1(C) and 1(D). See Methods and Materials section for further details on these spectral reproduction accuracy characterization experiments including comparison against ground truth spectra obtained using a conventional spectrometer.

In order to reconstruct the HD or full-frame image, we divided the full-frame diffractogram into overlapping patches. For each patch, a PSF was synthesized by linear interpolation of the PSFs from the three nearest calibration nodes. The patches were then individually reconstructed, and stitched together using a weighted sum with a tapered window function to reduce edge artifacts. Once the spectral image is generated, inferencing techniques are applied to extract scene-specific information.

## Experiments

We used a standard camera with a standard lens (focal length ~22 mm, f/5.6) and a monochrome CMOS sensor (IMX290LLR, Sony, 1945×1097 pixels, 2.9 μm pixel size). The diffractive filter array (DFA) was fabricated using nanoimprint lithography with pixel-width matching the sensor pixel width. The heights of the DFA pixels varied between 0 and 8.7 μm. The pattern was periodic over 8×8 DFA-pixels. The gap search space was 10 to 60 μm in 10 μm intervals. The wavelength domain (425–800 nm) was discretized into 25 bands, each 15 nm wide. The optimization determined the DFA pattern as shown in Fig. S2(A) and the DFA-image sensor gap was determined to be 50 μm. First, the pattern of the DFA was fabricated in a Silicon master using e-beam lithography and then transferred to the DFA polymer material using nanoimprint lithography. The refractive index of the DFA material is shown in Fig. S1. The cover glass of the image sensor was removed and the DFA was precisely aligned such that its XY axes matched those of the image sensor. A photograph of the DFA assembled on to the monochromatic image sensor is shown inset of Fig. S2(B). During imaging, the lens focus was adjusted using a high-resolution Siemens star chart, and both focus and f-number were fixed for all experiments. The camera was first validated by generating small spectral images (~122×122 pixels) of a narrowband (bandwidth = 15nm) beam illuminating a the white and colored squares of a Macbeth ColorChecker chart using narrowband illumination (bandwidth = 15 nm) at different wavelengths as well as broadband white light. An example result from each case is shown in Figs. 1(C) and 1(D), respectively, with full set of results in Fig. S6 and S7. We compared the reconstructed spectra to that from a conventional spectrometer (Jazz, Ocean Optics) and confirmed good agreement. We also generated the effective spectral responses of the camera, which confirmed spectral resolution between 30 and 50nm across the entire 440 to 800nm range (see Fig. S7). See Methods and Materials section for further details on these spectral reproduction accuracy characterization experiments. This experiment serves as an indirect way of measuring the spectral response of the camera. The measured spectral response shows good agreement with the simulated spectral response.

We demonstrated HD full-frame spectral imaging by first recording the diffractogram of a scene (illuminated by white LED light) spanning the field-of-view of the full-frame of the image sensor (Fig. 2(A)). Five of the 25 spectral channels are shown for clarity in Fig. 2(B-F). Other full-frame examples with comparison to ground truth obtained using spectrometers as well as full spectral data extraction and reconstructed synthetic RGB image generation are provided in Fig. S9, with further details of the experiment in the Methods and Materials section.

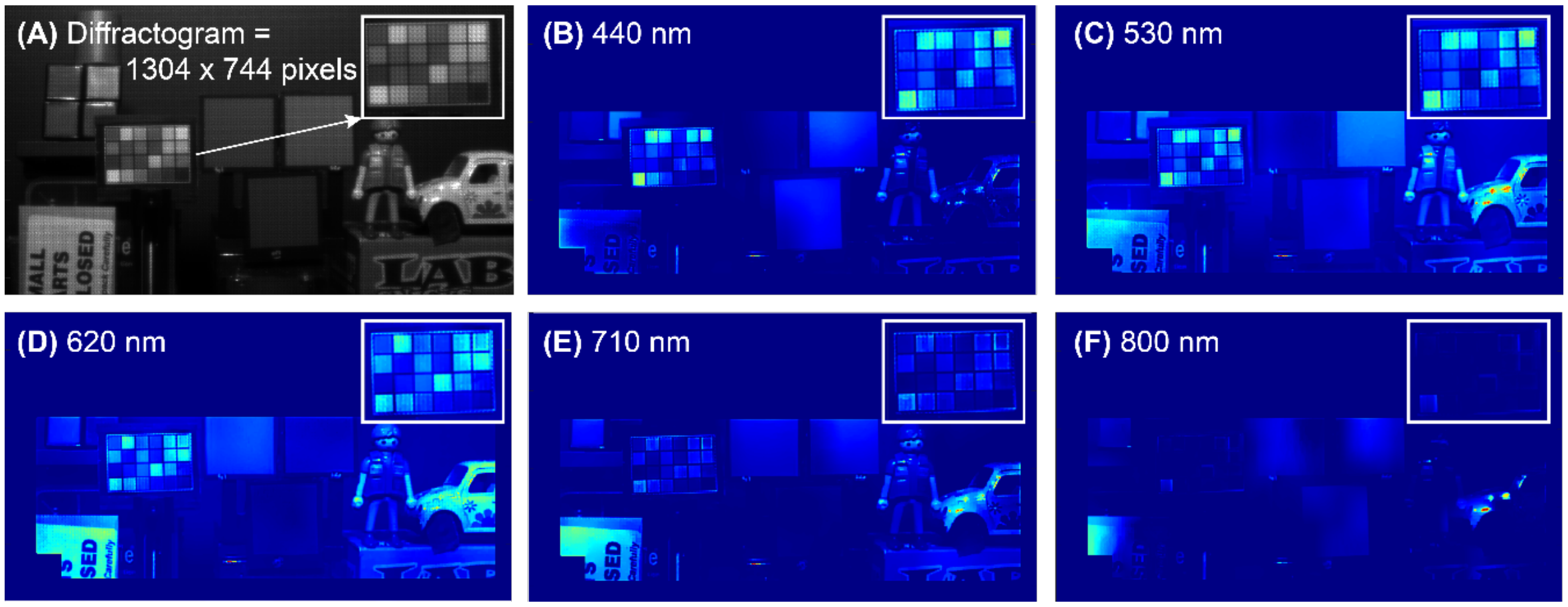

**Fig. 2: HD (~1MP) diffractive-spectral imaging**. **(A)** Diffractogram (1304 × 744 pixels ~ 1MP). Reconstructed spectral images: **(B-E)** 5 spectral channels out of 25 are shown for clarity. Top-right insets show magnified view of the Macbeth color chart. Different squares light up in different channels as expected. This scene is illuminated by white LEDs.

We validated our imager via proof-of-concept experiments in various domains. Despite the inherent reconstruction errors, our algorithm provides sufficient accuracy for meaningful inference from the image data cubes. We demonstrate its versatility with examples in tissue classification in biological samples, non-destructive food quality assessment, and stellar radiometry simulations for astronomical applications.

In our first experiment, we aimed to demonstrate the potential of our diffractographic approach for biological tissue imaging and identification. To achieve this, we obtained fresh chicken lungs and trachea and captured diffractogram images under consistent white LED illumination, avoiding pixel saturation. Utilizing our previously described algorithm, we reconstructed the spatio-spectral image cube, which we then transformed into RGB images for visual inspection (Fig. 3(A)). We manually segmented the images into lung and trachea regions to extract their spectral data. Fig. 3(B) shows different spectral slices highlighting how the two tissues absorb light differently at different wavelengths. The mean spectra for lung and trachea pixels, along with standard deviations, indicated unique spectral signatures due to varying absorption and scattering properties (Fig. 3(C)), which indicates the possibility of clear separability using the recovered spectra. To avoid overfitting our small dataset, we applied a Linear Discriminant Analysis (LDA) classifier to the labeled spectral data. The LDA effectively distinguished between the tissues, clearly separating them in a 2D component space (Fig. 3(D)). The RGB image of the reconstructed cube with classification results overlay (trachea in red, lungs in green) demonstrates accurate tissue classification (Fig. 3(A)).

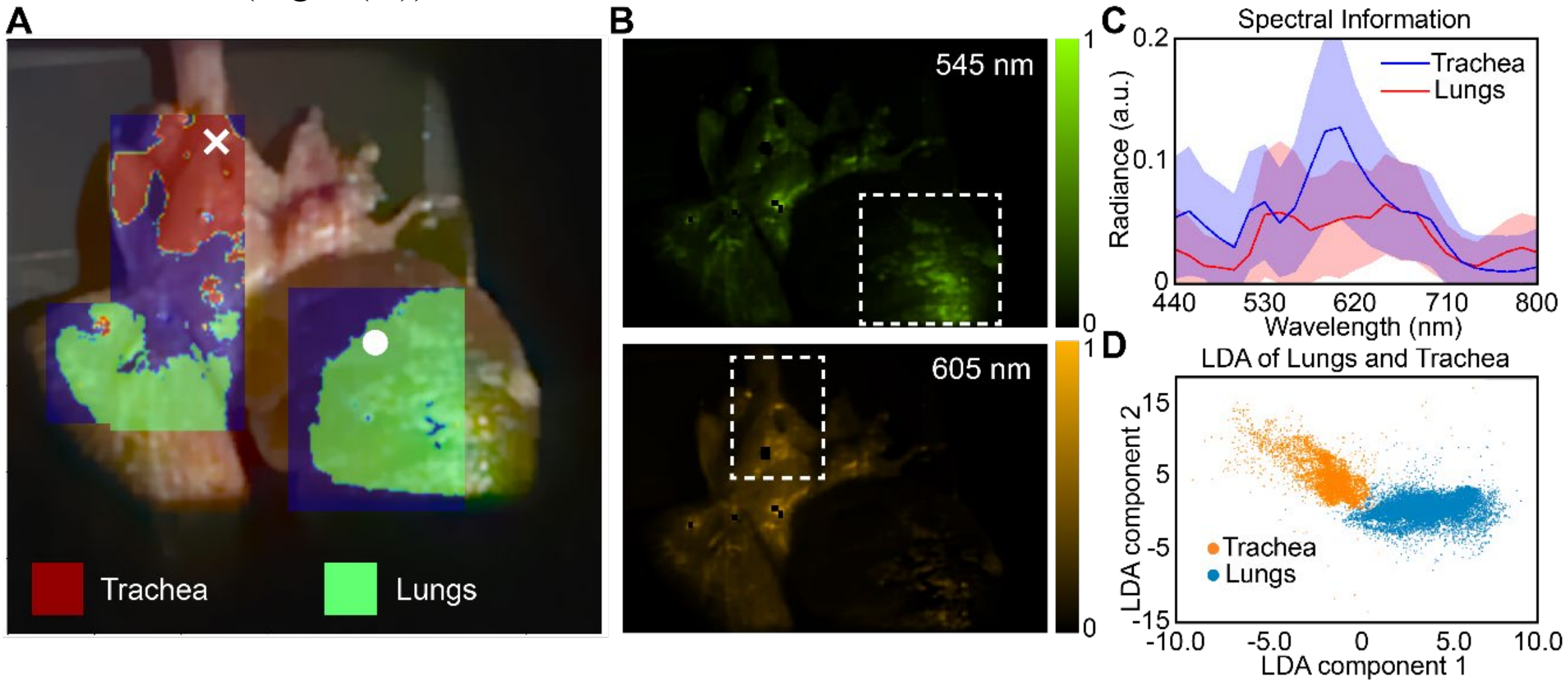

**Fig. 3: Tissue classification results. (A)** Reconstructed RGB image (344 x 344 pixels) of the chicken tissue with LDA classification overlays (lungs in green, trachea in red). **(B)** 2 spectral channels (2 out of 25 bands), indicating distinct spectral responses of the lungs and trachea. **(C)** Mean spectra (solid lines) and standard deviations (shaded areas) for lungs and trachea, highlighting their spectral differences. **(D)** LDA plot showing clear separation of trachea and lung tissues in a 2D component space.

The second experiment consisted of imaging 21 supermarket-bought strawberries for a span that covered 8 days. The collected diffractograms (Fig. 4(A)) were reconstructed to generate corresponding spectral images. The total dataset (346 images) was randomly split into 16 berries (265 images) for training and 5 berries (81 images) for testing. The spectra corresponding to fruit pixels were manually segmented, and their 25-band (the same 440–800nm, with 15nm bins) radiance spectra were used as the independent variable. Fig. 4(B) shows a visible spectral evolution over time, suggesting a functional relationship. Specifically, we observed a decrease in the green peak as the strawberries aged, consistent with previous studies *(53)*.

Despite anticipating a non-linear dependency, a multivariate linear regressor predicted the actual age of the fruit with considerable accuracy (Fig. 4(C)). As a secondary test, we trained a logistic classifier to predict if an observed strawberry was older than 5 days or not. The classification results can be seen in Fig. 4(D) as a receiver operating characteristic (ROC) plot. This curve illustrates the diagnostic ability of the binary classifier as its discrimination threshold varies, shows the trade-off between the true positive rate (sensitivity) and false positive rate (1-specificity). A diagonal ROC curve would indicate classification accuracy no better than random chance.

More sophisticated regressors could better model the non-linear relation between the fruit's spectral response and its aging, thereby improving accuracy in the regression and classification. However, the primary focus of this study is to demonstrate that diffractograms can capture, and the reconstruction can extract, the relevant latent spaces to perform robust and effective inference on this and similar applications, thereby showcasing the potential of our diffractographic approach in agricultural monitoring or similar problems.

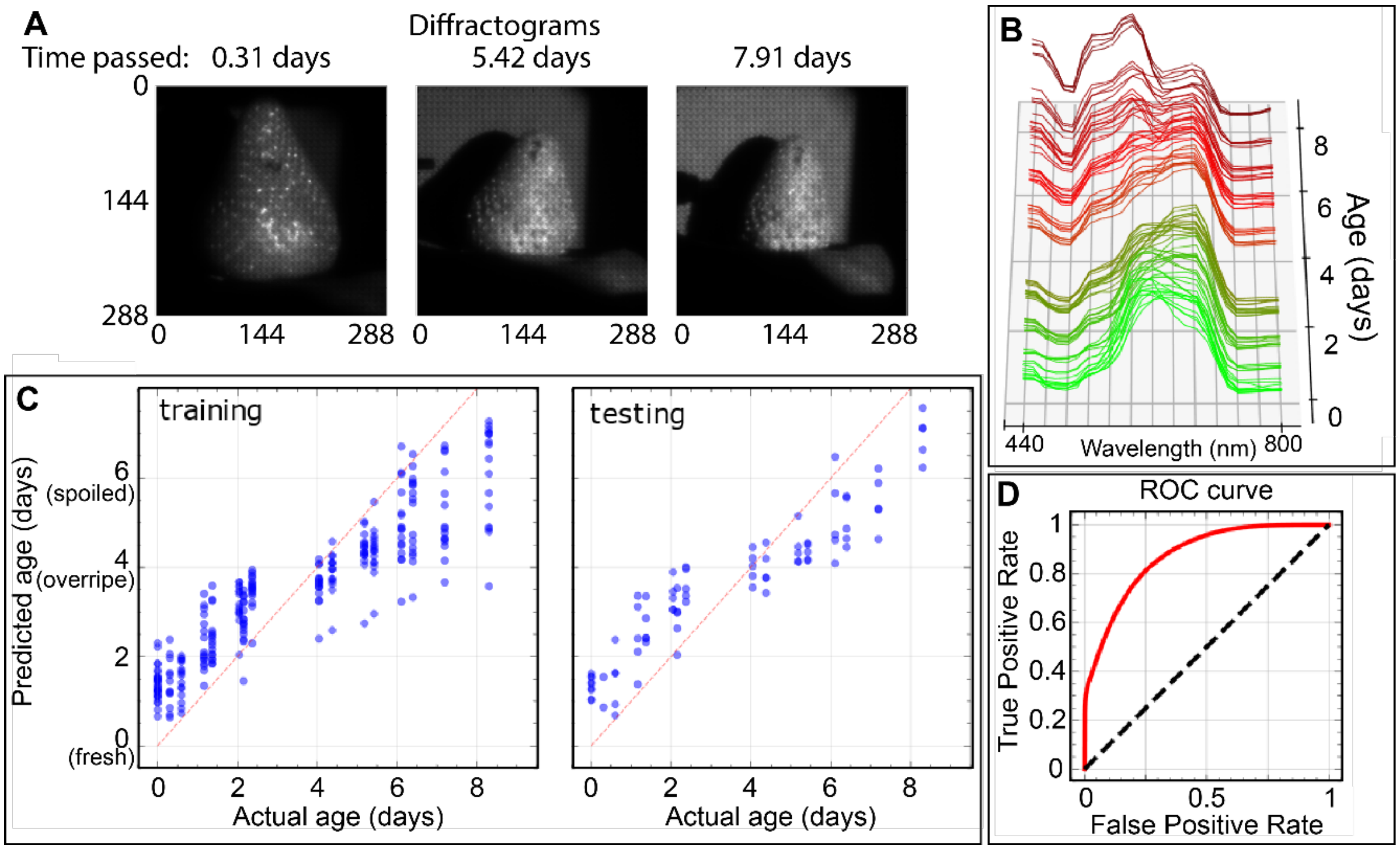

**Fig. 4: Predicting the age of a strawberry. (A)** Diffractogram images of strawberries in our training set at different times. **(B)** Mean radiance spectra from the reconstructed spectral images strawberries as a function of aging. **(C)** Predicted vs actual strawberry age using the training set (left) and test set (right) using a multivariate linear regressor. **(D)** ROC for the strawberry-age classification experiment.

Our third and final application used reconstructed spectra from diffractogram data to emulate the Johnson-Cousins (JC) UBVRI filter set in the Vis-NIR band whose letters correspond to ultraviolet, blue, visual, red and infrared filters, respectively. These filters facilitate standardized photometric measurements, enabling astronomers to study stellar properties, determine color indices, monitor variable stars, classify galaxies, and estimate redshifts *(54, 55)*. The system's widespread adoption ensures consistent and comparable data across different telescopes and instruments, making it integral to modern astronomical research *(56, 57)*.

In this example, we show that a diffractogram approach can be used to approximate the recovery of the signal using all filters, simultaneously. Instead of recovering the 25-bands as in previous experiments, we use a much smaller basis using the BVRI transmission curves, shown in Fig. S10. We do not use the ultraviolet (U) curve since it lacks meaningful support in our working spectral range. By integrating our PSF A with each JC transmission curve $C_{j\in\{B,V,R,I\}}(\lambda)$ with respect to wavelength, we obtain a dimensionally reduced, application-specific basis $B_j$:

$$B_j(u,v;x,y) = \int A(u,v;x,y,\lambda)C_j(\lambda)\,d\lambda, \quad j \in \{B,V,R,I\} \tag{3}$$

then the forward problem becomes

$$D(u,v) = \int \sum_{j\in\{B,V,R,I\}} S_j(x,y)B_j(u,v;x,y)\;dx\,dy \tag{4}$$

and the corresponding inverse problem to recover $S_{j\in BVRI}$ can be solved using the same methodology described earlier. Notably, being a smaller dimensional operator (4 vs 25 bands), the reconstruction can be done more efficiently.

The experiment consisted of simulating a stellar scene by back-illuminating a black opaque board with three pinholes using white LED light and added color filters (blue, green and red) to emulate spatially compact sources with broad spectra. The filters used for the left, middle, and right pinholes were green, red, and blue filters, respectively. After capturing the diffractograms, we reconstructed the 4-band (BVRI) spectral images. To obtain reference coefficients for comparison, we applied the JC transmission spectra to generate relative photometric measurements. The comparison between the recovered and reference coefficients is shown in Fig. 5(B).

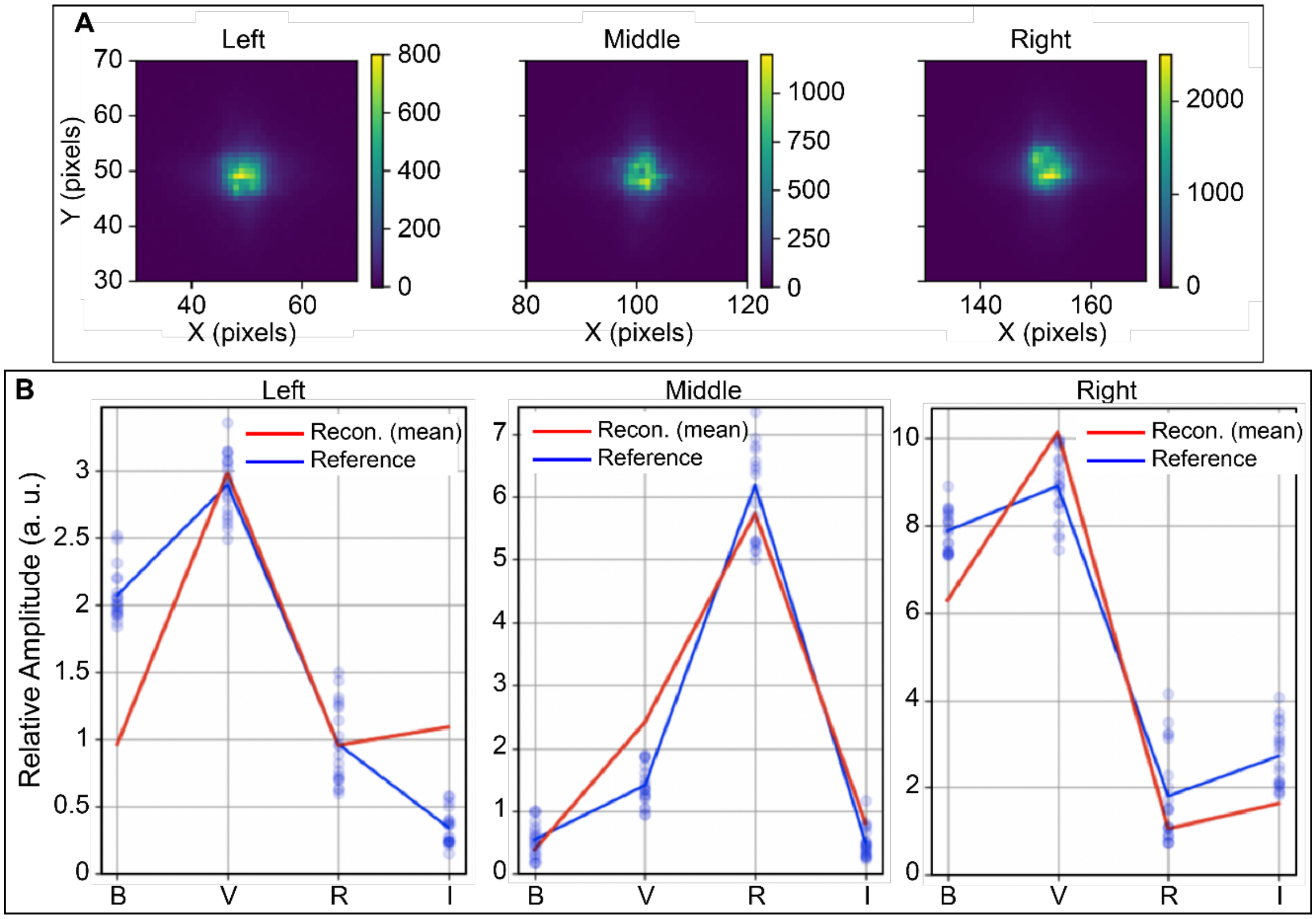

**Fig. 5: Predicting stellar filter responses. (A)** Raw diffractograms of the three simulated stars: left (green filter), middle (red filter), right (blue filter). Note that the relative intensity of the three stars is very different. **(B)** Reconstructed coefficients using the reduced basis comparing the ground truth (red line) and samples inside the simulated star pixels (blue dots). Vertical axes correspond to relative amplitudes (a.u.).

This experiment demonstrates that with a computational reconstruction approach, the basis can be chosen *a posteriori* and tailored to specific applications. In this instance, by reconstructing four distinct spectral images (BVRI) from a single diffractogram signal, we effectively bypass the need for traditional, lossy color filters. This not only maximizes light utilization but also eliminates the time-consuming process of sequential filter wheel rotations or mechanical scanning.

## Discussion

Our study demonstrates the transformative potential of using a diffractive filter array (DFA) for high-definition (HD) snapshot spectral imaging. By replacing the conventional Bayer filter in a standard camera with a transparent diffractive filter array, we can capture a single diffractogram that encodes both spatial and spectral information. This diffractogram can then be computationally decoded to reconstruct a spectral image cube with 25 spectral bands in the 440–800 nm range.

We underscore that hyperspectral image cubes are rarely the end product; they are used for inference tasks such as classification, segmentation, and anomaly detection. It is also well-known that hyperspectral datasets exhibit high redundancy, with strong correlations between spectral bands. And most importantly, the features of interest often lie on a low-dimensional manifold embedded within the high-dimensional hyperspectral dataset. This observation challenges the notion that a higher number of spectral bands is always beneficial, as for many inference tasks, low-dimensional approximations are not only sufficient but can also be more efficient.

Our approach offers several characteristics that can be advantageous in certain circumstances. First, it enables snapshot capability, overcoming the problems with scan-and-stitch methods. Second, the diffractogram itself can be viewed as a form of optical compression, efficiently storing spatio-spectral content in a compact, information-rich 2D array. This is particularly beneficial for applications with very limited storage or transmission bandwidth, such as in airborne or satellite imaging scenarios. This is even more pronounced in the case of time-varying situations. For instance, a raw uncompressed hyperspectral video with a resolution of 512×512 pixels and 100 bands (16-bit per voxel) and 1000 frames would require 50 GB of storage, while the equivalent 1-band (16-bit per pixel) diffractographic video would need just 500 MB.

Third, the flexibility to perform reconstructions offline and on-demand after data capture is valuable in scenarios where on-board computational resources are limited. Fourth, traditional hyperspectral imaging systems discretize spectra *a priori* into hundreds of bands, which can be inefficient and wasteful when some applications only require a few functionals of the spectral curves to perform inference tasks. In contrast, our diffractogram encodes spectral information continuously, allowing for *a posteriori* selection of an application-specific spectral basis. Not only can it yield smaller image cubes, but it also translates into faster and more stable reconstructions. In our Johnson-Cousins filter experiment, we demonstrated the ability to reconstruct a reduced set of four bands, precisely tailored to the needs of the application, avoiding the unnecessary processing of large data cubes. Finally, unlike other spectral imaging methods that rely on lossy filters, our DFA is highly transparent, ensuring maximum utilization of incident light. It also has the potential for low-cost production and miniaturization of the imaging system.

The proof-of-concept experiments in tissue classification, food quality assessment, and synthetic stellar imaging demonstrate that, despite the loss of information in the diffractogram projection and errors introduced during the reconstruction process, the method can successfully extract relevant information to perform robust and accurate inference tasks. This suggests that there exists a class of problems for which the unique characteristics and advantages of the diffractogram imaging system are well-suited.

## Materials and Methods

### PSF calibration

During PSF calibration, a pseudo-ideal point source was generated at narrowband wavelengths (bandwidth = 15 nm) and stepped across one 8×8 pixel unit cell of the DFA at 10 different locations in the field-of-view. A super-continuum laser source (SuperK EXTREME, NKT Photonics) coupled to a tunable filter (SuperK VARIA, NKT Photonics) served as the illumination source. The laser beam was expanded and collimated (Fig. S3) to approximate 3" diameter to make it uniform in intensity across the beam profile. A detailed schematic diagram of the setup is provided in Fig. S3(A, D). This beam was then focused to generate a point source using an objective lens (Olympus RMS20X, Thorlabs) and a 300 μm pinhole (P300D, Thorlabs). The objective-pinhole setup was mounted on a motorized micrometer-actuated platform for precise XY movement. The tunable filter selected 25 wavelengths (440-800 nm, at 15 nm bandwidth). For each wavelength, the point source was moved across the 8×8 pixel unit cell, with camera exposure times adjusted to avoid pixel saturation. The diffractogram at each location-wavelength combination was averaged over 8 frames to reduce noise (Fig. S3). Thus, we are able to record $A(u, v; x, y, \lambda)$, the spatio-spectral point spread function (PSF). Fig. S3(B, C) shows exemplary point spread function (PSF) values recorded during the calibration step. Exemplary values from the full calibration data cube for 5 equally spaced wavelengths and five locations on the X-Y grid are shown in Fig. S3(B) with magnified views of three of these in Fig. S3(C).

Since we empirically determined that the PSF remains consistent for a distance up to 200 pixels from where it is measured, therefore, in order to attain HD or full-frame imaging (1304×744 spatial pixels, × 25 spectral bands), we measured the PSF at ten different locations across the sensor frame as shown in Fig. S4. The schematic of performing the additional sparse calibrations is shown in Fig. S4(A). The map of the calibration points on the full frame is shown in Fig. S4(B). We performed a total of 10 calibrations, following the same method as described earlier. For these additional calibrations, the camera was physically moved in its plane in the X and Y axis, but keeping the distance from the calibration/image plane the same (550 mm) as shown in Fig. S4(A). This moves the PSF location in the sensor plane as shown in Fig. S4(B). The point $C_1$ in Fig. S4(B) corresponds to the first "on-axis" calibration for which, the results are shown in Figs. 1(A-B), S6 and S7. The sparse additional calibrations at the locations $C_2$, $C_3$, $C_4$, $C_5$, $C_6$, $C_7$, $C_8$, $C_9$, $C_{10}$ correspond to 34%, 57%, 90% field of view (FOV) in the X axis, 45% FOV and 93% FOV in the diagonal direction and the four corners of the entire FOV, respectively. For imaging characterization, we filled the full frame field-of-view at the object plane with various toys, colored paper, etc. and illuminated the scene using white LED flashlights.

### Spectral reproduction accuracy characterization

We first characterized the response of the DFA-enabled camera to narrowband illumination. We cut out the white square from an X-Rite ColorChecker Classic (MSCCC) color chart and placed it at the object plane. The chart contains 24 color squares (each approximately 40 × 40 mm) on a 205.7 × 289 mm board. We used square 19, as shown in Fig. S5. The supercontinuum source was directed onto this square using mirrors, creating an elliptical illumination due to oblique incidence (Fig. S6(A)). We selected 25 wavelengths from 440 to 800 nm with a 15 nm bandwidth using a tunable filter and captured the corresponding diffractograms, adjusting camera exposure to avoid saturation. Dark frames were recorded with the laser off. Spectral images were reconstructed following the manner described previously. Then, we sampled five 8×8 points: one at the center and four with a ±10 pixel offset in each quadrant, averaging the values across wavelengths to extract representative spectra corresponding to each image. Ground truth was obtained using a spectrometer (Jazz, Ocean Optics) in reflection mode, capturing the spectrum via an 8 μm single-mode fiber. The computationally reconstructed synthetic RGB images for the individual wavelengths are shown in Fig. S6(B). The comparison of the spectral response of our DFA-enabled camera compared to ground truth spectra obtained using the spectrometer for each wavelength, in shown in Fig. S6(C) and shows good agreement at the peak for each wavelength. The result corresponding to $\lambda$ = 530 nm is presented as a summary in Fig. 1(C). Full spectral data cubes can be extracted for all the wavelengths and are shown for three wavelengths at 470, 530, and 620 nm, representative of blue, green and red colors in Fig. S6(D). Comparison to ground truth spectra obtained using the conventional spectrometer and measurement of the full-width-at-half-maximum (FWHM) of the spectral resolution are provided in Fig. S7.

Next, we characterized the spectral reproduction performance of the DFA-enabled camera to broadband illumination. We used colored squares from the Macbeth ColorChecker Classic (X-Rite). Each square was placed at the object plane, illuminated by a collimated beam from a white LED flashlight. The beam was collimated through multiple reflections over a ~1.57 m path, creating uniform illumination on the object. An iris limited the beam's extent, resulting in elliptical illumination due to the oblique illumination, as shown in the schematic in Fig. S8(A). We imaged 18 colored and 2 grayscale squares from the Macbeth chart. Fig. 1(D) shows summarized results from square 14, with complete results for all other squares in Fig. S8. Images were captured with 32 frames averaged to reduce noise, and dark frames captured similarly. The averaged dark frame was subtracted from the averaged image frame to obtain the diffractogram. Ground truth was captured using a spectrometer in reflection mode. Full spectral data cubes can be extracted and are shown for three squares numbered 8, 14, and 15, representative of blue, green and red colors, in Fig. S8(D).

## HD spectral imaging

For the full-frame HD spectral imaging experiment, we imaged a number of plastic toys, colored papers, color squares cut from Macbeth chart, etc. and flood illuminated the scene using a white LED flashlight (ClipLight LED flashlight). The imaging and reconstruction procedures are same as described before. A schematic of the experimental setup is shown in Fig. S9(A). The DFA-enabled camera records a diffractogram of the scene in snapshot mode. A synthetic RGB image of the scene is shown in Fig. S9(B). The dimensions of this image are 1144×548 pixels. The red dots indicate locations of PSF measurement (Fig. S4) whereas the white check marks indicate locations where we extract and compare the spectra to ground truth obtained using the spectrometer, as shown in Fig. S9(E). Fig. S9(C) shows the full spectral data at wavelengths 440-800 nm. The hyperspectral data cube shown in Fig. S9(D) is constructed from the data in Fig. S9(C), with spatial $(x, y)$ and spectral $(\lambda)$ data arranged along orthogonal axes. Two slices (purple, dashed and green dashed) show $x\lambda$ and $y\lambda$ planes, revealing different spectral responses of different materials of the objects comprising the scene. Images were captured with 32 frames averaged to reduce noise, and dark frames captured similarly. The averaged dark frame was subtracted from the averaged image frame to obtain the diffractogram. Ground truth was captured using a spectrometer in reflection mode.

### Chicken tissue preparation and imaging

We were supplied fresh chicken tissue from a local butcher by Dr. Chakravarthy Reddy. The tissue was extracted from the animal and imaged within 4 hours of extraction to retain freshness. It was propped up against a black foam board in order to be illuminated in a vertical position facing the camera. The imaging setup is the same as illustrated in Fig. S9, with the tissue taking the position of the objects. The same white LED flashlight was used to illuminate the scene.

### Synthetic stellar imaging

In order to simulate a stellar scene, we punched three pinholes in a completely black opaque board and covered them with three color filters (blue, green and red color). We illuminated this from the back using a white LED flashlight. Thus, this essentially generates three points at different colors located at the image plane. We used our camera to record the diffractogram of this scene in snapshot mode. The reference was recorded using a commercial hyperspectral camera (Ultrix X20, Cubert). A schematic of the opical setup is shown in Fig. S11.

**Acknowledgments**

We are grateful to Dr. Chakravarthy B. Reddy, Huntsman Cancer Institute for assistance with the tissue imaging experiment and valuable discussions regarding the results. We also acknowledge Prof. Henry I. Smith, Massachusetts Institute of Technology and Prof. Berardi Sensale-Rodriguez for valuable feedback on the manuscript.

**Funding:**
Office of Naval Research (N6560-NV-ONR)

**Author contributions:**
Conceptualization: AM, FGC, FGV, RM
Methodology/Experiments: AM, MM, FGC, FGV
Synthetic stellar imaging experiments: AM, FGC, SNQ, FS
Visualizations: AM, FGC
Supervision: AM, RM
Writing—original draft: AM, FGC, RM
Writing—review & editing: AM, FGC, FGV, RM

**Competing interests:** FGC, FGV and RM are part of Lumos Inc. All other authors declare they have no competing interests.

**Data and materials availability:** All data are available in the main text or the supplementary materials.

Supplementary Materials for

# HD snapshot diffractive spectral imaging and inferencing

Apratim Majumder, Monjurul Meem, Fernando Gonzalez del Cueto, Fernando GuevaraVasquez, and Rajesh Menon

*Corresponding author. Email: apratim.majumder@utah.edu

**This PDF file includes:**

Figs. S1 to S10

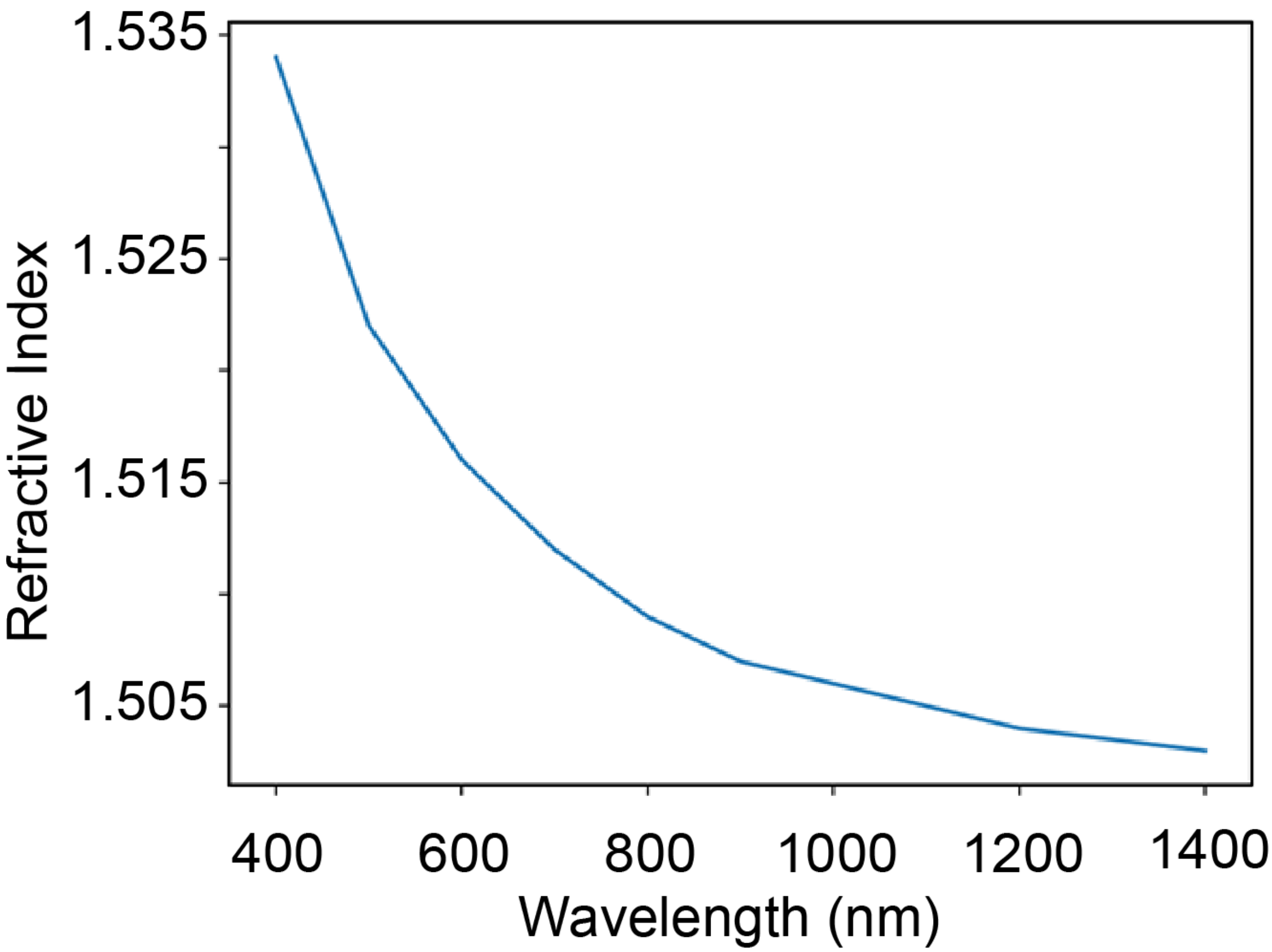

**Fig. S1. Refractive index of the diffractive filter array (DFA) material.**

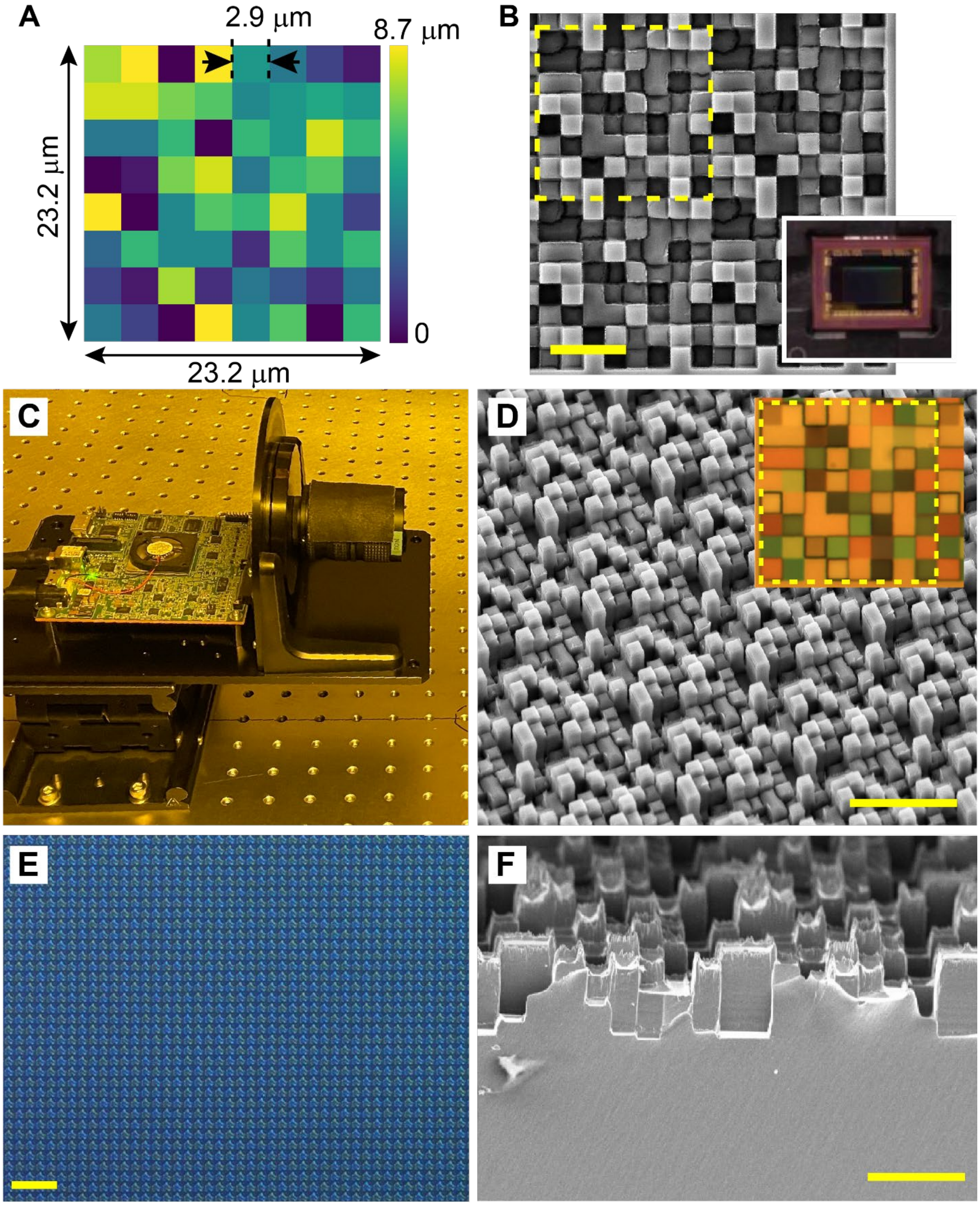

**Fig. S2. DFA-enabled camera assembly. (A)** Optimized height distribution of one unit-cell of the DFA, composed of an 8×8 square pixel array, where each pixel has a side length of 2.9 µm. **(B)** Scanning electron micrograph of the Silicon master of the DFA (scale bar = 10 µm). One unit cell has been indicated by the yellow dashed square. Inset shows the image sensor chip after the DFA was replicated in polymer using nanoimprint lithography and assembled on to the monochrome image sensor post cover glass removal. **(C)** Photograph of the assembled camera. **(D)** Tilted scanning electron micrograph of the DFA master fabricated in Silicon. Scale bar = 20 μm. Inset shows optical micrograph of one unit cell of the DFA indicated by the yellow dashed square with dimension of 23.2 μm × 23.2 μm. **(E)** Optical micrograph of the DFA replicated in the DFA material using nanoimprint lithography. Scale bar = 1 mm. **(F)** Cross-sectional scanning electron micrograph of the DFA after replication. Scale bar = 10 μm.

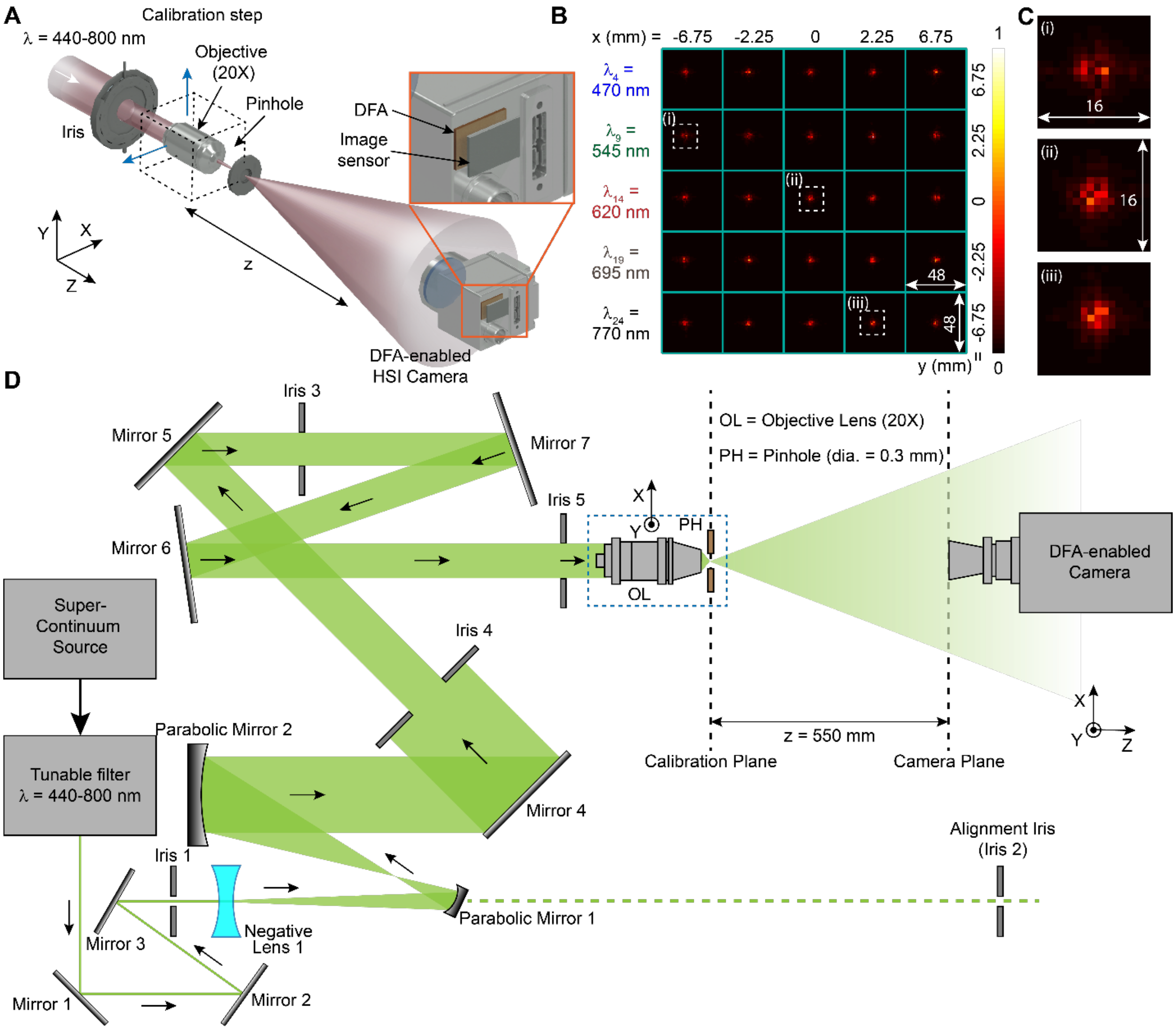

**Fig. S3. Diffractive Filter Array (DFA) PSF calibration. (A)** Schematic diagram showing the optical setup used for the calibration step. The illumination (λ = 400-800 nm) is from a broadband supercontinuum source (SuperK EXTREME, NKT Photonics) that is coupled to a tunable filter (SuperK VARIA, NKT Photonics) to select the individual wavelengths and set the wavelength center bandwidths. The beam from the laser source is expanded, collimated and incident on the objective lens + pinhole assembly. The blue arrows indicate the directions of movement of the objective lens + pinhole assembly in the XY plane. The distance between the camera and the objective lens + pinhole assembly, z = 550 mm. The objective lens + pinhole assembly is stepped in the XY plane over an 8×8 step matrix corresponding to one unit cell of the DFA, while the illumination wavelength is changed to 25 discrete values, evenly spaced between 440 and 800 nm with 15 nm bandwidths for each. This allows us to record the PSF calibration matrix $\boldsymbol{A}(\boldsymbol{u},\boldsymbol{v};\boldsymbol{x},\boldsymbol{y},\boldsymbol{\lambda})$. **(B)** Raw calibration data for 5 exemplary equally spaced wavelengths and 5 locations on the XY plane. Each cropped frame has 48×48 pixels. **(C)** Magnified (16×16 pixels) views of three points (i-iii) indicated in (B). **(D)** Detailed schematic of the optical setup (top-view) showing all optical components used. Mirrors 1-3: PF10-03-P01, Thorlabs, Mirrors 4-7: PF20-03-P01, Thorlabs, Negative Lens 1: ACN254-040-A, Thorlabs, Parabolic Mirror 1 (CM127-010-P01,

Thorlabs) and Parabolic Mirror 2 (CM750-500-P01, Thorlabs) comprise the beam expander. Negative Lens 1 is used to initially expand the beam from a diameter of 5 mm to 12.5 mm. The final diameter of the collimated beam past Mirror 4 is 3 inches. Pinhole PH: P300D, Thorlabs, diameter = 300 µm. Objective Lens OL: Olympus RMS20X, Thorlabs. The objective lens focuses the collimated beam to a point at its working distance and the pinhole is collocated at this plane to spatially filter the beam and produce a pseudo-ideal point source. The objective + pinhole combination is mounted on a motorized micrometer actuated 3-axis stage (not shown here) to move it in XY plane. The camera sits on a lab jack for coarse vertical alignment.

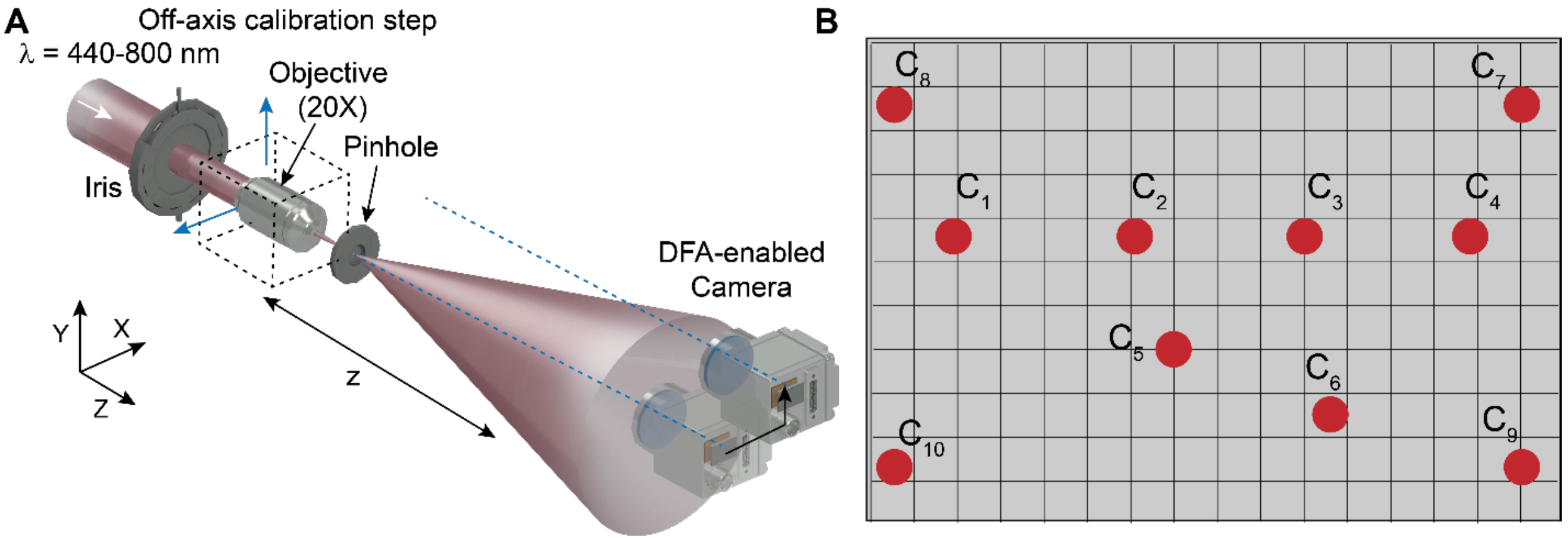

**Fig. S4. Additional calibration map for full-frame HD snapshot spectral reconstructions. (A)** Schematic of performing the additional sparse calibrations. The camera was physically moved in its plane in the X and Y axis, but keeping the distance from the calibration/image plane the same (550 mm). This moves the PSF location in the image sensor plane. We performed a total of 10 calibrations. **(B)** Map of the calibration points on the full frame. The point C1 corresponds to the first "on-axis" calibration. The sparse additional calibrations at the locations C2, C3, C4, C5, C6, C7, C8, C9, C10 correspond to 34%, 57%, 90% field of view (FOV) in the X axis, 45% FOV and 93% FOV in the diagonal direction and the four corners of the entire FOV, respectively.

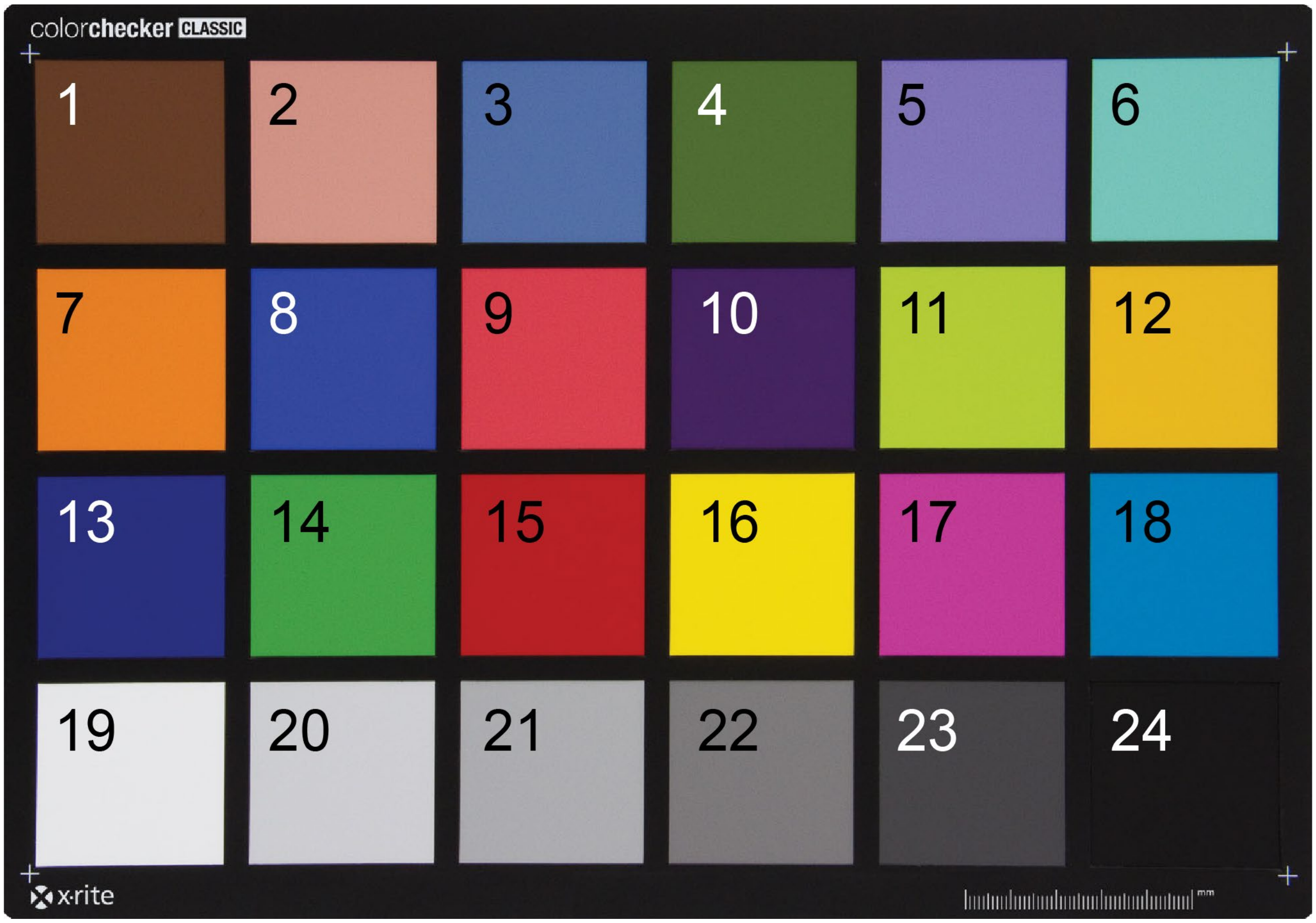

**Fig. S5. Index for ColorChecker chart.** X-Rite ColorChecker Classic (MSCCC) color chart with numbered indexes for the colors. These numbers have been referenced in the experiments.

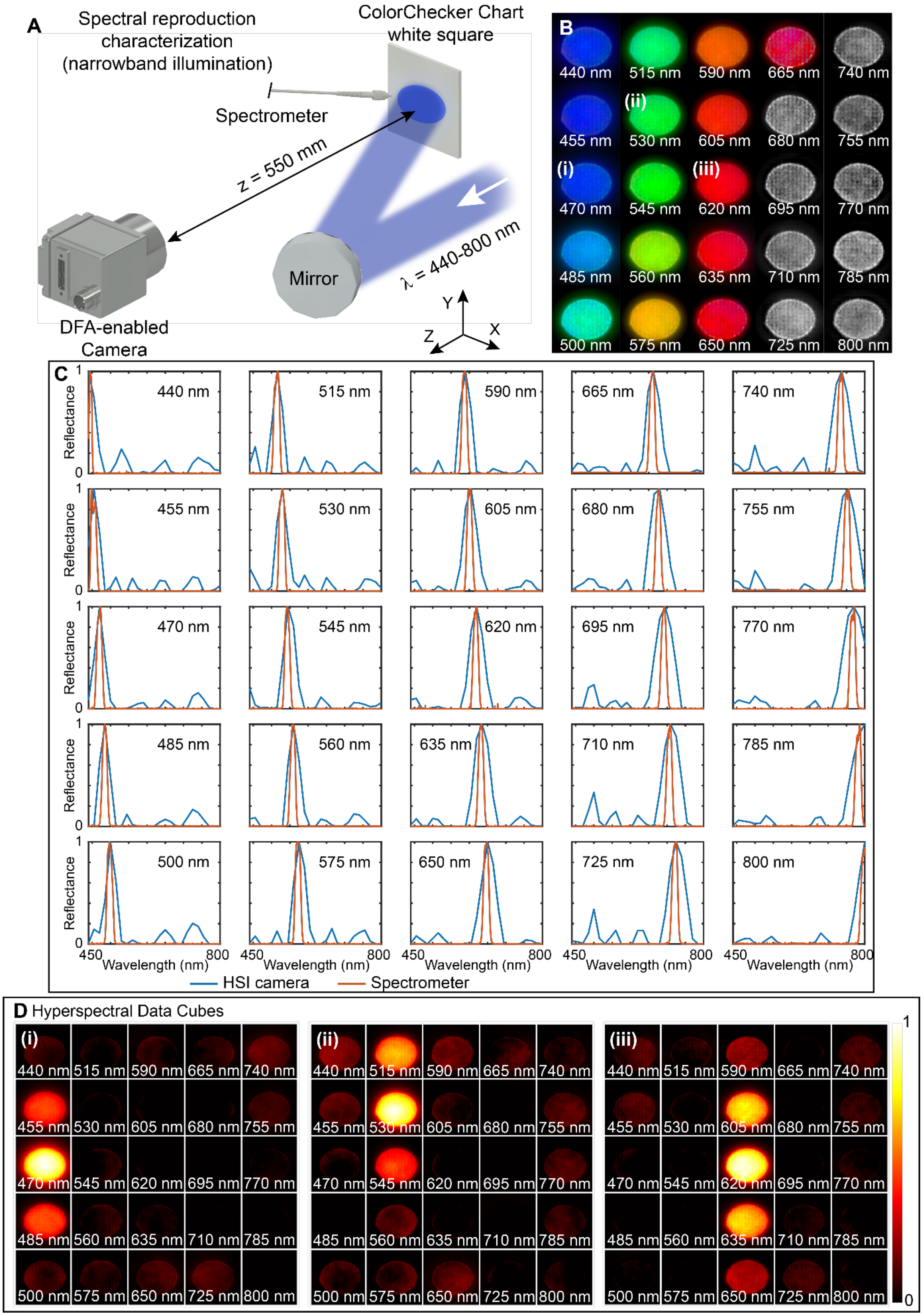
A
Spectral reproduction characterization (narrowband illumination)
ColorChecker Chart white square
Spectrometer
z = 550 mm
Mirror
λ = 440-800 nm
DFA-enabled Camera
B
C
Reflectance
Wavelength (nm)
HSI camera
Spectrometer
D Hyperspectral Data Cubes

**Fig. S6. Characterization of spectral reproduction accuracy to narrowband illumination. (A)** Schematic diagram showing the optical setup used for the DFA spectral response characterization experiment **to narrowband illumination**. The illumination (λ = 400-800 nm) from the supercontinuum source (SuperK EXTREME, NKT Photonics + SuperK VARIA tunable filter) is expanded, collimated and then guided by mirrors on to a white square cut out from a standard X-Rite ColorChecker Classic (MSCCC) color chart (also known as a Macbeth Color chart). We can select the individual wavelengths and set their bandwidths using the SuperK VARIA. 25 wavelengths, equally spaced between 440 and 800 nm, with individual bandwidths of 15 nm were selected. The incident beam, reflected at an angle from the mirror onto the white Macbeth chart square forms an elliptical spot and serves as the object for this experiment. The image of this object is captured by our DFA-enabled camera. A spectrometer is also used to obtain the ground truth of the spectra. **(B)** Computationally reconstructed synthetic RGB images of the individual wavelengths. Wavelengths at and greater than 680 nm are represented in grayscale as there is no color information there. The dimension of each image is 133×133 pixels, and the dimensions of the elliptical area of illumination are major axis ~ 98 pixels and minor axis ~ 80 pixels. **(C)** Spectral reproduction of our DFA-enabled camera compared to ground truth spectra obtained using the spectrometer for each wavelength, showing good agreement at the peak for each wavelength. **(D)** Full spectral data cubes shown for three wavelengths at (i) 470, (ii) 530, and (iii) 620 nm, representative of blue, green and red colors.

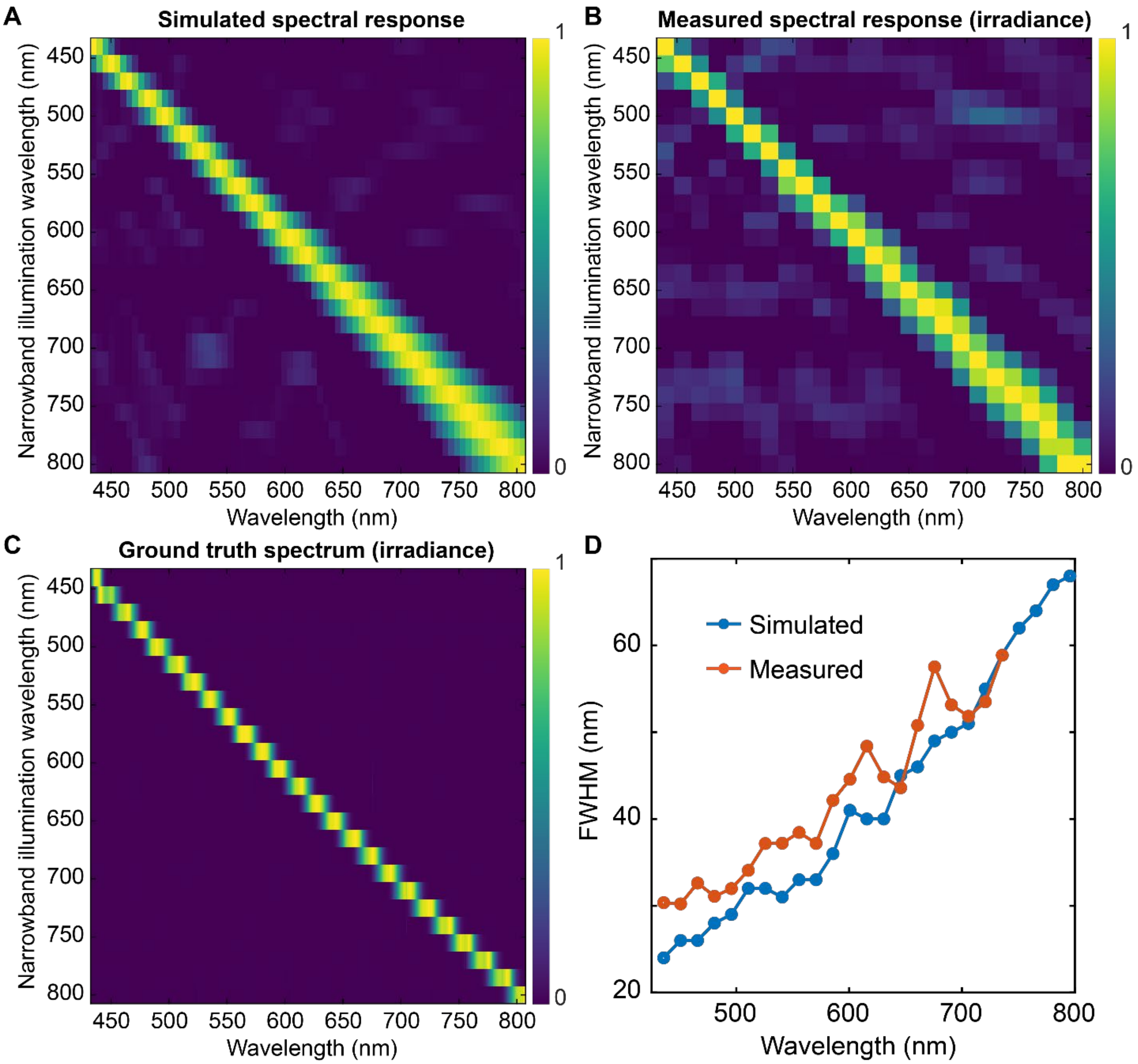

**Fig. S7. Spectral reproduction accuracy to narrowband illumination. (A)** Simulated narrowband spectral response of the DFA, shown for 25 equally spaced wavelengths with peaks from 440 to 800 nm. **(D)** Measured spectral response of the DFA for 25 equally spaced wavelengths with center peaks from 440 to 800 nm. This is a heatmap generated from the data presented in Fig. S6(C), blue curves. **(E)** Ground truth spectra measured using the conventional spectrometer. This is a heatmap generated from the data presented in Fig. S6(C), orange curves. **(F)** Simulated and measured spectral full-width at half maximum (FWHM) of the narrowband spectra in (A) and (B).

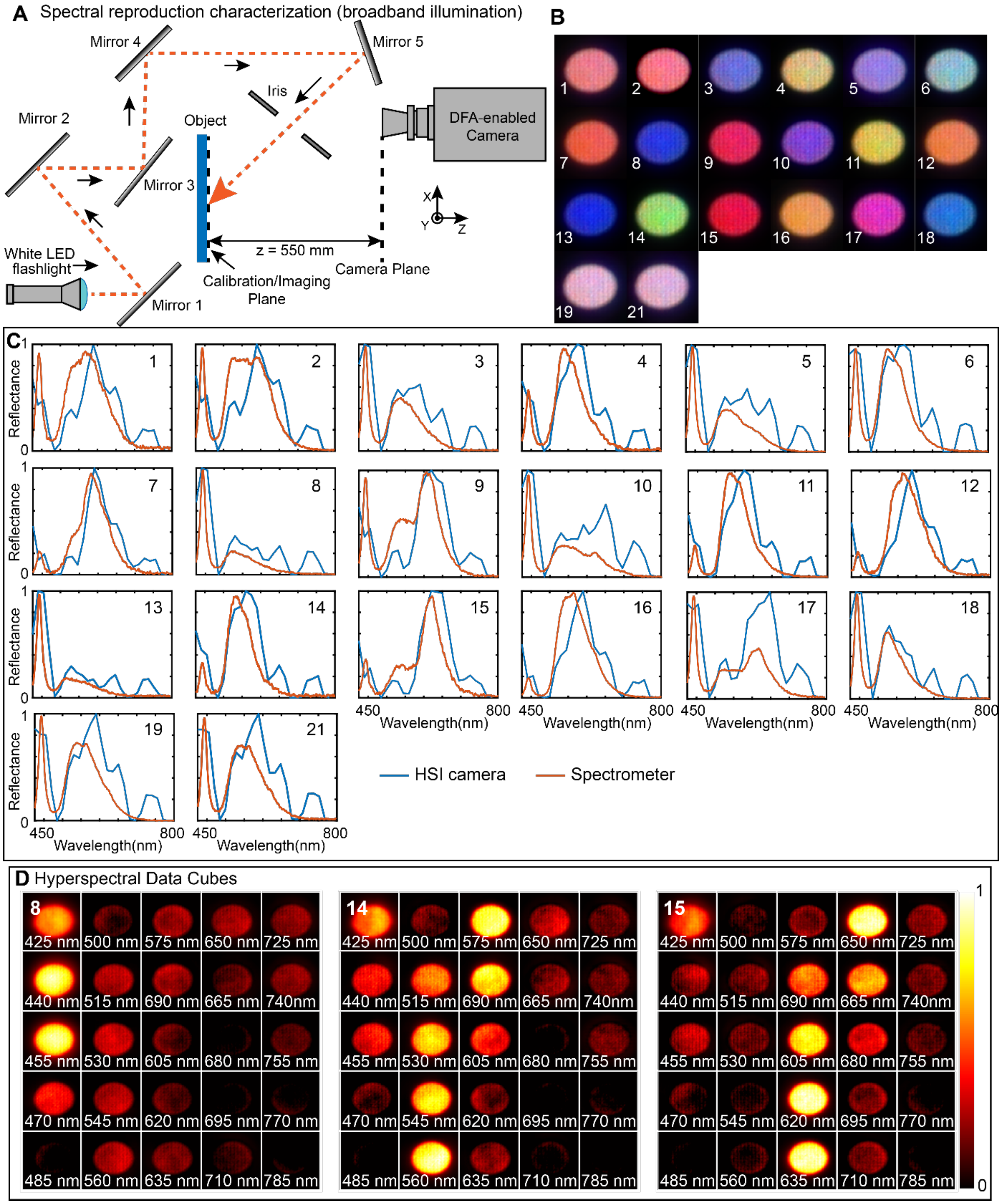

**Fig. S8. Characterization of spectral reproduction accuracy to broadband illumination. (A)** Schematic diagram showing the optical setup. The illumination from a white LED flashlight (Maglite Pro LED) was collimated using multiple reflections between flat mirrors and ultimately guided onto a colored square cut out from the Macbeth ColorChecker chart. The illumination extent was limited using an iris in the path of the beam. Being an oblique incidence illumination, the spot shape was elliptical on the colored square. This serves as the object for this experiment.

The image of this object is captured by our DFA-enabled camera. A spectrometer was also used to obtain the ground truth of the spectra in reflection mode off the Macbeth chart square. **(B)** Computationally reconstructed synthetic RGB images of the individual colored squares with numbered indexes corresponding to Fig. S5. The dimension of each image is 122×122 pixels, and the dimensions of the elliptical area of illumination are major axis ~ 87 pixels and minor axis ~ 74 pixels. **(C)** Spectral reproduction of our DFA-enabled camera compared to ground truth spectra obtained using the spectrometer for each colored square. **(D)** Full spectral data cubes shown for three squares numbered 8, 14, and 15, representative of blue, green and red colors.

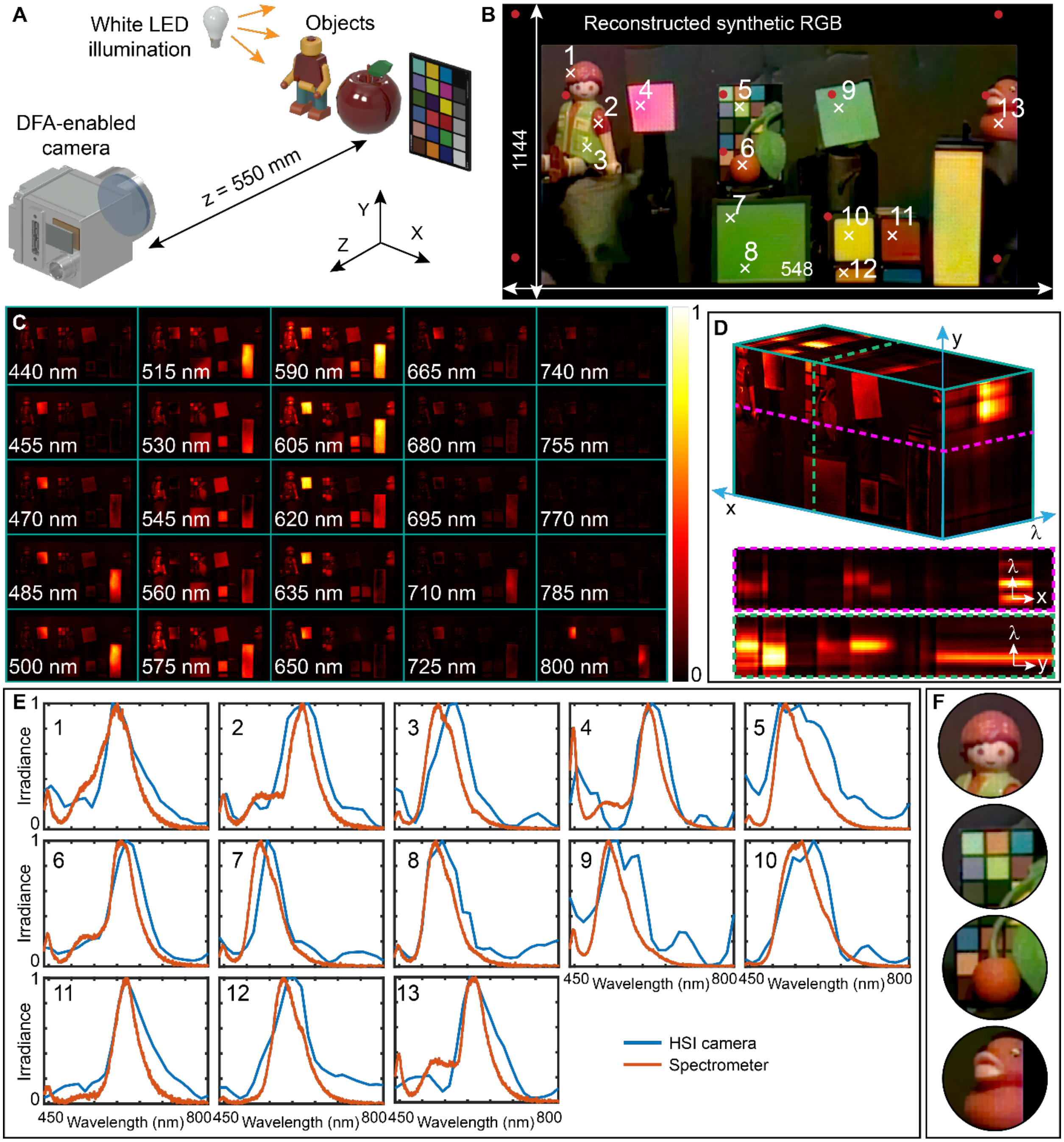

**Fig. S9. Full-frame snapshot spectral imaging. (A)** Schematic diagram showing the setup used for the imaging experiment. A scene comprising colorful objects of different materials such as plastic (toys), paper, rubber (duck), etc. are flood lit by incoherent illumination from a white LED flashlight (ClipLight LED flashlight). The camera records the diffractogram of the scene that can then be used to reconstruct the full spectral data cube, generate synthetic RGB image as well as extract spectral image of any location in the scene. **(B)** Synthetic RGB image from reconstructed spectral data cube showing excellent color fidelity. The red dots indicate locations were the PSF measurements were performed. The white checks are locations where we extracted the spectra from the spectral data and compared against ground truth obtained by the spectrometer. **(C)** Full

spectral data of the scene. **(D)** Spectral data cube constructed from (C). Two slices indicated by the green and purple dashes are extracted to show the data in xλ and yλ planes, revealing different spectral responses of different materials of the objects comprising the scene. **(E)** Spectral comparison between data obtained by our camera (blue curve) with that obtained using the spectrometer (orange curve) for the white check mark locations indicated in (B). **(F)** Magnified views of some of the objects in the scene from the RGB image showing image detail.

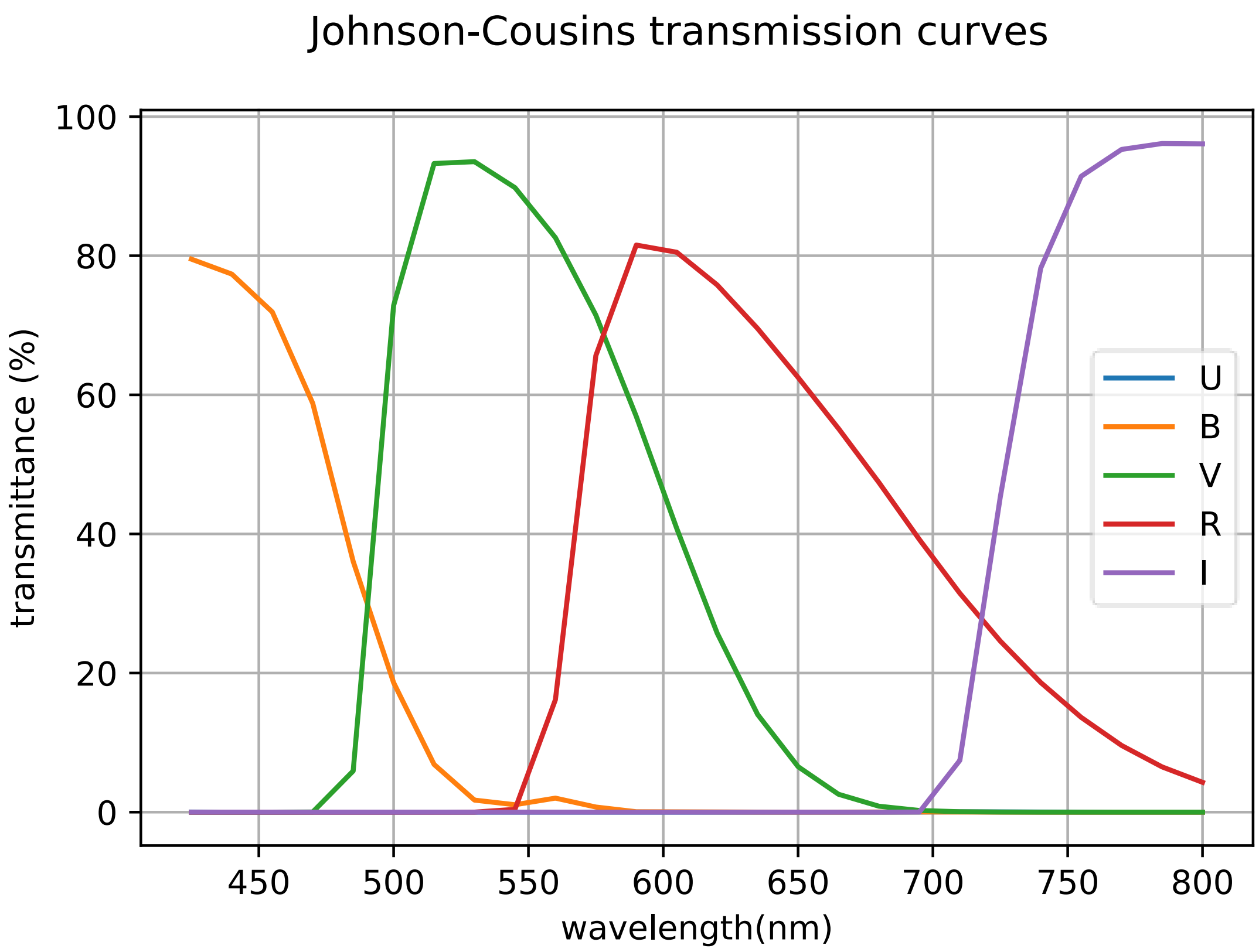

**Fig. S10. Johnson-Cousins spectral curves**: The five curves in the spectral range and discretization of our system, 25 uniform bands in the 440–800nm. The ultraviolet filter (U) does not have any support in this range so it is not used in our experiment.

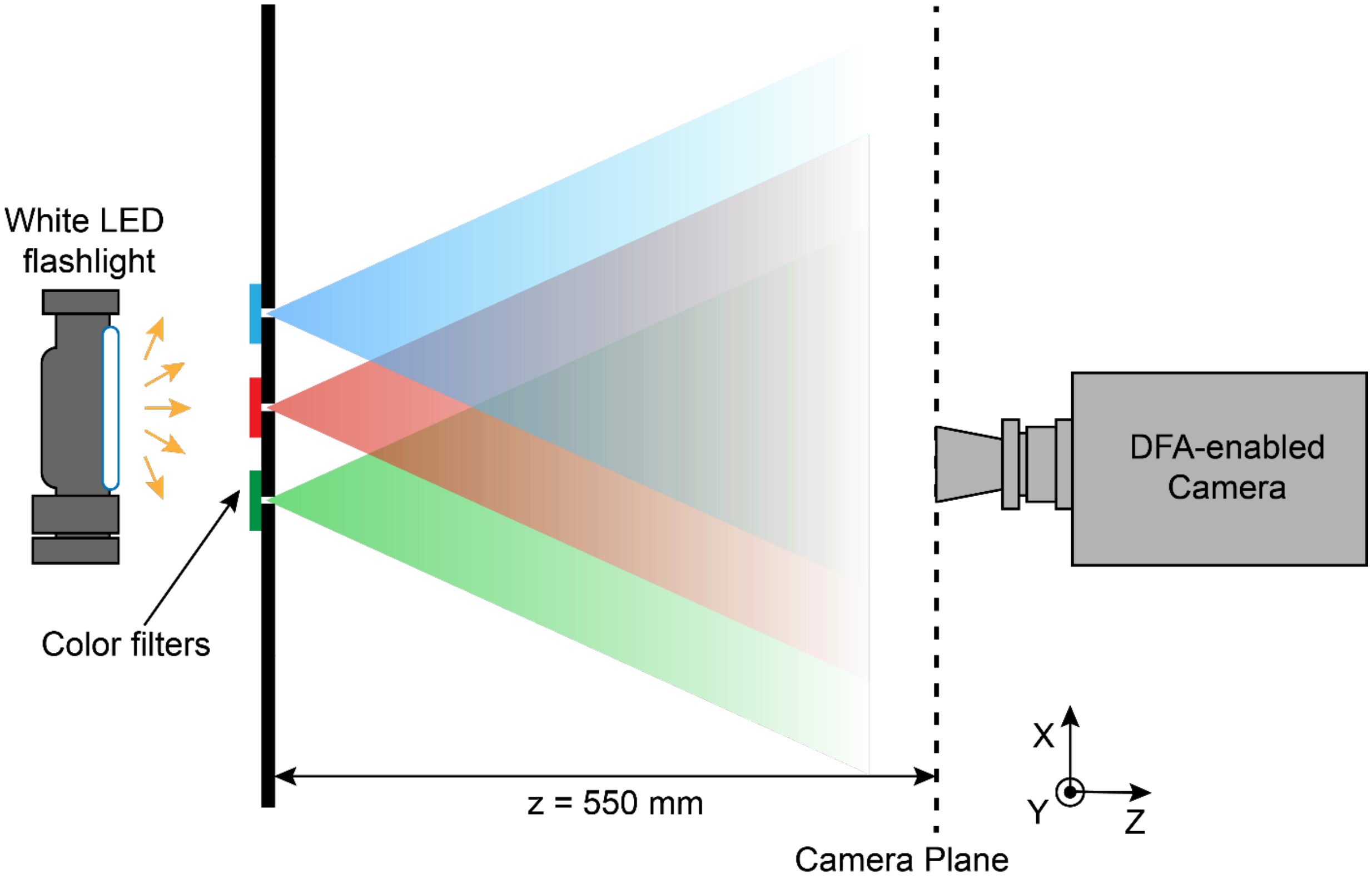

**Fig. S11. Synthetic stellar imaging**: Schematic showing the optical setup used for performing the synthetic stellar imaging experiments. We punched three pinholes in a completely black opaque board and covered them with three color filters (blue, green and red color). We illuminated this from the back using a white LED flashlight. We used our camera to record the diffractogram of this scene in snapshot mode. The reference was recorded using a commercial hyperspectral camera (Ultrix X20, Cubert), placed at the location of our camera.